\input harvmac
 
%%%%%%%%%%%%%%%%%%%%%%%%%%%%%%%%%%%%%%%%%%%%%%%%%%%%%%%%%%%%%%%%%%%
%%%  modify title page
%%%%%%%%%%%%%%%%%%%%%%%%%%%%%%%%%%%%%%%%%%%%%%%%%%%%%%%%%%%%%%%%%%%
\def\Title#1#2{\rightline{#1}\ifx\answ\bigans\nopagenumbers\pageno0
\vskip0.5in
\else\pageno1\vskip.5in\fi \centerline{\titlefont #2}\vskip .3in}

%\def\listrefs{\footatend\bigskip\bigskip\immediate\closeout\rfile
%\writestoppt \baselineskip =13pt\centerline{{\secfont References}}
%\bigskip{\frenchspacing\parindent =20pt \escapechar +'
%\input\jobname.refs \vfill\eject}\nonfrenchspacing} 
%%%%%%%%%%%%%%%%%%%%%%%%%%%%%%%%%%%%%%%%%%%%%%%%%%%%%%%%%%%%%%%%%%%%%%%%%%%%

\noblackbox
\parskip=1.5mm
%\def\semi{;~}

%%%%%%%%%%%%%%%%%%%%%%%%%%%%%%%%%%%%%%%%%%%%%%%%%%%%%%%%%%%%%%%%%%%%%
  
\def\npb#1#2#3{{\it Nucl. Phys.} {\bf B#1} (#2) #3 }
\def\plb#1#2#3{{\it Phys. Lett.} {\bf B#1} (#2) #3 }
\def\prd#1#2#3{{\it Phys. Rev. } {\bf D#1} (#2) #3 }
\def\prl#1#2#3{{\it Phys. Rev. Lett.} {\bf #1} (#2) #3 }

\def\jmp#1#2#3{{\it J. Math. Phys.} {\bf #1} (#2) #3 }
\def\cmp#1#2#3{{\it Commun. Math. Phys.} {\bf #1} (#2) #3 }
\def\pr#1#2#3{{\it Phys. Rev. } {\bf #1} (#2) #3 }

\def\bb#1{{\tt hep-th/#1}}

\def\rmp#1#2#3{{\it Rev. Mod. Phys.} {\bf #1} (#2) #3 }
\def\jetp#1#2#3{{\it Sov. Phys. JEPT} {\bf #1} (#2) #3 }

\def\app#1#2#3{{\it Astropart. Phys. } {\bf #1} (#2) #3 }
\def\ap#1#2#3{{\it Adv. Phys. } {\bf #1} (#2) #3 }
\def\pr#1#2#3{{\it Phys. Rep. } {\bf #1} (#2) #3 }
\def\aop#1#2#3{{\it Ann. Phys. (NY)} {\bf #1} (#2) #3 }
\def\nat#1#2#3{{\it Nature } {\bf #1} (#2) #3 }
\def\cqg#1#2#3{{\it Class. Quant. Grav. } {\bf #1} (#2) #3 }
\def\grg#1#2#3{{\it Gen. Rel. Grav. } {\bf #1} (#2) #3 }

%%%%%%%%%%%%%%%%%%%%%%%%%%%%%%%%%%%%%%%%%%%%%%%%%%%%%%%%%%%%%%%%%%%%%
%%%%%%%%%%%%%%%%%%%%    some definitions    %%%%%%%%%%%%%%%%%%%%%%%%%
%%%%%%%%%%%%%%%%%%%%%%%%%%%%%%%%%%%%%%%%%%%%%%%%%%%%%%%%%%%%%%%%%%%%%

%%%%%%%%%%%%%%%%%%%%%%%%%%%%%%%%%%%%%%%%%%%%%%%%%%%%%%%%%%%%%%%%%%%%%

\def\dj{\hbox{d\kern-0.347em \vrule width 0.3em height 1.252ex depth
-1.21ex \kern 0.051em}}

\def\tr{{\rm tr\,}}

%%%%%%%%%%%%%%%%%%%%%%%%%%%%%%%%%%%%%%%%%%%%%%%%%%%%%%%%%%%%%%%%%%%%%%
%%%%%%%%%%%%%%%%%%%%%%%        references         %%%%%%%%%%%%%%%%%%%%%%
%%%%%%%%%%%%%%%%%%%%%%%%%%%%%%%%%%%%%%%%%%%%%%%%%%%%%%%%%%%%%%%%%%%%%%%
\lref\rsc{E. \'Alvarez, \prd{31}{1985}{418\semi}
R. Brandenberger and C. Vafa, \npb{316}{1988}{391\semi} 
A. Tseytlin and C. Vafa, \npb{372}{1992}{443\semi}
C.R. Nappi and E. Witten, \plb{293}{1992}{309\semi}
M. Gasperini and G. Veneziano, \app{1}{1993}{317\semi}
R. Brustein and P.J. Steinhardt, \plb{302}{1993}{196\semi}
M.A.R. Osorio and M.A. V\'azquez-Mozo, \plb{320}{1994}{259\semi}
E. Kiritsis and C. Kounnas, \plb{331}{1994}{51\semi}
E.J. Copeland, A. Lahiri and D. Wands, \prd{50}{1994}{4868\semi}
R. Brustein and G. Veneziano, \plb{329}{1994}{429\semi}
N.A. Batakis, \plb{353}{1995}{450\semi}
N.A. Batakis and A.A. Kehagias, \npb{449}{1995}{248\semi}
T. Banks, M. Berkooz, S.H. Shenker and G. Moore, \prd{52}{1995}{3548\semi}
M. Gasperini, M. Giovannini and G. Veneziano, \prd{52}{1995}{6651\semi}
J.D. Barrow and M.P. D\c{a}browski, \prd{55}{1997}{630.}}
\lref\rbarrow{J.D. Barrow and K. Kunze, 
{\it Inhomogeneous String Cosmologies,} (\bb{9701085}.)}
\lref\rven{G. Veneziano, {\it Inhomogeneous Pre-Big Bang Cosmology}, 
Preprint CERN-TH-97-042
(\bb{9703150}).}
\lref\rps{R. Poppe and S. Schwager, \plb{393}{1997}{51.}}
\lref\rnpst{J. Polchinski, \rmp{68}{1996}{1245.}}
\lref\rwitt{E. Witten, \npb{443}{1995}{85.}}
\lref\rpenrose{R. Penrose, in 
{\it General Relativity: An Einstein Centenary Survey,} 
ed. S.W. Hawking and W. Israel, Cambridge 1979.}
\lref\rmisner{C.W. Misner, \prl{22}{1969}{1071.}}
\lref\rguth{A. Guth, \prd{23}{1981}{347.}} 
\lref\rland{as told by I.M. Khalatnikov.}
\lref\rblk{V.A. Belinskii, I.M. Khalatnikov and 
E.M. Lifshitz, \ap{31}{1982}{693} (and references therein).}
\lref\rcchm{M. Carmeli, Ch. Charach and S. Malin, \pr{76}{1981}{79.}}
\lref\rcchf{M. Carmeli, Ch. Charach and A. Feinstein, \aop{150}{1983}{392.}}
\lref\rverd{E. Verdaguer, \pr{229}{1993}{1.}}
\lref\rszekerez{P. Szekeres, \nat{228}{1970}{1183.}}
\lref\rkhapenr{K.A. Khan and R.Penrose, \nat{229}{1971}{185.}}
\lref\rfi{A. Feinstein and J. Ib\'a\~nez, \prd{39}{1989}{470.}}
\lref\rbk{V.A. Belinskii and I.M. Khalatnikov, \jetp{30}{1970}{1174;}
\jetp{32}{1971}{169.}}
\lref\rgeroch{R. Geroch, \jmp{13}{1972}{394.}}
\lref\rkz{V.A. Belinskii and V.E. Zakharov, \jetp{48}{1978}{985.}}
\lref\hfm{Z. Hassan, A. Feinstein and V. Manko, \cqg{7}{1990}{L109.}}
\lref\rkit{C. Hoenselaers, W. Kinnersley and B. Xanthopoulos, 
\jmp{20}{1979}{2530\semi}
D.W. Kitchingham, \cqg{1}{1984}{677.}}
\lref\rehlers{J. Ehlers, Ph.D. Dissertation, Hamburg 1957.}
\lref\rlwch{P.S. Letelier, \jmp{20}{1979}{2078\semi} 
J. Waiwright, W. Ince and B. Marshman, \grg{10}{1979}{259\semi}
Ch. Charach and S. Malin, \prd{19}{1979}{1058.}}
\lref\rgowdy{R.H. Gowdy, \prl{27}{1976}{826.}}
\lref\rtt{R. Tabensky and A.H. Taub, \cmp{29}{1973}{61.}}
\lref\buscher{T. Buscher, \plb{159}{1985}{127;} \plb{194}{1987}{59;} 
\plb{201}{1988}{466.}}
\lref\rschw{C. Hull and P. Townsend, \npb{438}{1995}{109\semi}
J.H. Schwarz, \plb{360}{1995}{13.}}
\lref\rmah{J. Maharana, 
{\it S-duality and compactification of the type-IIB superstring action}, 
Preprint Fermilab-97-046-T (\bb{9703009}).}
\lref\rhe{S.W. Hawking and G.F.R. Ellis, {\it The Large Scale 
Structure of Space-Time},
Cambrigde 1973.}
\lref\raabl{G.T. Horowitz and D.L. Welch, \prl{71}{1993}{328\semi}
E. \'Alvarez, L. Alvarez-Gaum\'e, J.L.F. Barb\'on and Y. Lozano, 
\npb{415}{1994}{71.}}

%%%%%%%%%TEXT%%%%%%%%%%%%%%%%%%%%%%%%%%%%%%%%%%%%%%%%%%%%%%%%%%%%%%%%
%%%%%%%%%%%%%%%%%%%%%%%%%%%%%%%%%%%%%%%%%%%%%%%%%%%%%%%%%%%%%%%%%%%%%
%%%%%%%%%%%%%%%%%%          title page       %%%%%%%%%%%%%%%%%%%%%%%%
%%%%%%%%%%%%%%%%%%%%%%%%%%%%%%%%%%%%%%%%%%%%%%%%%%%%%%%%%%%%%%%%%%%%%

\line{\hfill EHU-FT/9703}  
\line{\hfill {\tt hep-th/9704173}}
\vskip 1.5cm

\Title{\vbox{\baselineskip 12pt\hbox{}
 }}
{\vbox {\centerline{Closed Inhomogeneous String Cosmologies}
}}

\vskip 0.8cm

\centerline{$\quad$ {A. Feinstein\foot{wtpfexxa@lg.ehu.es}, 
Ruth Lazkoz\foot{wtblasar@lg.ehu.es} and 
M.A. V\'azquez-Mozo\foot{wtbvamom@lg.ehu.es}
 }}

\medskip

\centerline{{\sl Dpto. de F\'{\i}sica Te\'orica}}
\centerline{{\sl Universidad del Pa\'{\i}s Vasco}}
\centerline{{\sl Apdo. 644, E-48080 Bilbao, Spain}}

 \vskip 1.8cm

\noindent
We present a general algorithm which permits to construct solutions 
in string cosmology for heterotic and type-IIB superstrings in four
dimensions. Using a chain of transformations applied
in sequence: conformal, T-duality and $SL(2,{\bf R})$ rotations, 
along with the usual generating techniques associated to Geroch  
transformations in Einstein frame, we obtain  solutions with 
all relevant low-energy remnants of the string theory. To exemplify our
algorithm we present an inhomogeneous string cosmology 
with $S^{3}$ topology of spatial sections, discuss some  
properties of the solution and point out some subtleties involved in
the concept of homogeneity and isotropy in string cosmology.

%%%%%%%%%%%%%%%%%%%%%%%%%%%%%%%%%%%%%%%%%%%%%%%%%%%%%%%%%%%%%%%%%%%%%%

\Date{04/97}
%\draft

%%%%%%%%%%%%%%%%%%%%%%%%%%%%%%%%%%%%%%%%%%%%%%%%%%%%%%%%%%%%%%%%%%%%%%%%%%
%%%%%%%%%%%%                text begins                        %%%%%%%%%%%
%%%%%%%%%%%%%%%%%%%%%%%%%%%%%%%%%%%%%%%%%%%%%%%%%%%%%%%%%%%%%%%%%%%%%%%%%%

\newsec{Introduction}

Cosmological implications of string theory are receiving quite a broad 
attention
these days (for a necessarily incomplete list of recent and 
not-so-recent references 
see \refs\rsc\refs\rps\refs\rbarrow\refs\rven). This is not by chance
since, given the lack of traces of stringy effects in ``low energy'' 
particle physics, the 
cosmological scenario appears as a perfect arena for the search of 
string blueprints 
in the early Universe. 

In spite of the impressive progress in the study of string theory 
beyond the perturbative
regime (for a review see, for example, \refs\rnpst), we still 
lack a full-fledged non-perturbative formulation which allows a 
description of the early Universe at Planck time, where the classical 
concepts of space and time cannot be used any longer. In the meantime, 
one is bound to study classical cosmology using the low-energy effective 
action induced by string theory, which generalizes General Relativity by 
including other massless fields. As a consequence, in string cosmology 
one continues to ask essentially the same questions earlier formulated 
in  the framework of Einstein Relativity, but now posed in the presence 
of those extra degrees of freedom. The central puzzle of  
theoretical cosmology remains as well the same, and is concerned 
mainly with the question of why the present Universe looks as 
isotropic and homogeneous at large scales. The desire to answer this 
last question leads one to consider the 
initial conditions and the evolution
of the Universe at the beginning of expansion\foot{We will not 
dwell here on different approaches and views as to how to solve 
this problem, but just mention some central ideas such as 
chaotic cosmology program \refs\rmisner,
C-gravitational entropy \refs\rpenrose\ and the inflationary
scenario \refs\rguth.}. 

Apart from a few exceptions \refs\rbarrow\refs\rven, most 
of the work in string
cosmology is focused on homogeneous solutions. However, since
in General Relativity the generic solution near the 
cosmological singularity is neither isotropic nor homogeneous, one is 
led to study cosmological models with less symmetry, leaving aside not only 
the principle of isotropy but homogeneity as well. 
The ideas to study such a general 
behavior were  inspired by Landau \refs\rland\ and  
worked out in details in  classical papers by 
Belinski, Lifshitz and Khalatnikov\foot{In string cosmology 
the BLK approach has been recently discussed in the context 
of the Pre-Big Bang scenario by Veneziano 
\refs\rven.} (BLK) \refs\rblk. The techniques due to BLK are 
strongly rooted in physical intuition, but lack however rigor when 
approximations
are used in a highly non-linear regime. Although the conclusions of this
analysis are probably correct at large, one must necessarily confront
them against the analytic behavior of exact solutions.

Because of the mathematical complexity of the generic inhomogeneous models,
one usually deals with solutions in which homogeneity is broken in 
only one direction. 
The space-times obtained this way are usually referred to as $G_{2}$ or
Einstein-Rosen cosmologies \refs\rcchm \refs\rcchf \refs\rverd\ and are 
thought to
provide the leading approximation to a general solution near the 
initial singularity \refs\rbk. It is important to mention here 
the relation of the Einstein-Rosen
cosmologies to the interesting problem of scattering of plane waves in 
General Relativity \refs\rkhapenr \refs\rszekerez. This relation, 
considered first in \refs\rfi, consists just in inverting the arrow of 
time of the vacuum, or vacuum with massless fields, inhomogeneous 
solutions. After all,  the approach to the focusing two-surface in 
the collision of self-gravitating plane waves 
(not necessarily gravitational)  is just the time inverse of the  
behavior near the singularity in a $G_2$ cosmology. This being so, 
the study of the behavior of string induced $G_2$ cosmologies may 
lead to a better understanding of such a highly
non-linear process as plane wave scattering. 
	
The main purpose of this paper is to describe a sufficiently general
algorithm which permits the construction of new 
solutions of string cosmology starting from vacuum solutions of the 
Einstein field equations. Vacuum $G_2$
space-times will be used as the basic building blocks of our string solutions.
Before going any further in the description of our generating technique,
it will be useful to briefly mention some facts about the low-energy effective
theory of the different superstring models and their symmetries. 

\subsec{String low-energy effective actions}

On general grounds, the massless bosonic
sector of superstrings includes the gravitational field $G_{\mu\nu}$, 
the dilaton $\phi$, with vacuum expectation value determining the string 
coupling constant, and the antisymmetric rank-two tensor $B_{\mu\nu}^{(1)}$, 
besides other fields depending on the particular superstring model 
under study. The lowest order effective
action for the massless fields can be written as
\eqn\univ{
S_{\rm eff}={1\over (\alpha^{'})^{D-2\over 2}}
\int d^{D}x\sqrt{G}e^{-2\phi}\left[R+4(\partial\phi)^{2}-
{1\over 12}(H^{(1)})^{2}\right]+S_{\rm md},
}
where $H^{(1)}=dB^{(1)}$ is the field strength 
associated with the NS-NS two-form
and $S_{\rm md}$ is a model-dependent part which includes other massless 
degrees of freedom. When $D<10$ some of these massless fields correspond
to gauge and moduli fields associated with the specific compactification 
chosen, and the dilaton $\phi$ appearing in \univ\ is related to the 
ten-dimensional
dilaton $\phi_{10}$ by $2\phi=2\phi_{10}-\log V_{10-D}$, where $V_{10-D}$ 
is the volume of the
internal manifold measured in units of $\sqrt{\alpha^{'}}$.

We restrict our attention to generic degrees of freedom, leaving aside the 
internal components of the ten-dimensional fields.
In the heterotic string case, $S_{\rm md}$ 
contains the Yang-Mills action for the 
background gauge fields $A_\mu^{a}$, which we set to zero in the following.
In the case of the model-dependent part of the type-IIB superstring 
things are slightly more involved; among the
massless degrees of freedom in the R-R sector we find, 
along with a pseudo-scalar
$\chi$ (the axion) and a rank-two antisymmetric tensor $B_{\mu\nu}^{(2)}$, 
a rank four self-dual form $A_{\mu\nu\sigma\lambda}^{\rm sd}$. The 
presence of this self-dual form spoils the covariance of the effective action  
for the massless fields,
since there is no way of imposing  the self-duality condition in a generally 
covariant
way. But, if we set $A^{\rm sd}$ to zero, we can write a covariant action for 
the remaining fields with
$$
S^{\rm IIB}_{\rm md}=-{1\over (\alpha^{'})^{D-2\over 2}}
\int d^{D}x\sqrt{G}\left[{1\over 2}(\partial\chi)^{2}+
{1\over 12}(H^{(1)}\chi+H^{(2)})^{2}\right]
$$
Notice that the R-R fields do not couple directly
to the dilaton. Thus, the lower dimensional ($D<10$) R-R fields $\chi$
and $B^{(2)}_{\mu\nu}$ are obtained from the ten-dimensional ones through
$\chi=\sqrt{V_{10-D}}\chi_{10}$ and $B^{(2)}=\sqrt{V_{10-D}}B_{10}^{(2)}$.

By combining the dilaton and the axion of the ten-dimensional type-IIB 
superstring into 
a single complex field $\lambda=\chi_{10}+ie^{-\phi_{10}}$, it is 
possible to check that the 
bosonic effective action is invariant under the $SL(2,{\bf R})$ 
transformation\foot{
The two antisymmetric tensors $B^{(1)}_{\mu\nu}$ and $B^{(2)}_{\mu\nu}$ 
also transform
as a doublet under $SL(2,{\bf R})$. Notice that here we are using 
the string frame in
which the metric is not invariant.}
$\lambda\rightarrow (a\lambda+b)/(c\lambda+d)$, $G_{\mu\nu}\rightarrow
|c\lambda+d|G_{\mu\nu}$ \refs\rschw. Writing this transformation in 
terms of four-dimensional fields ($D=4$) we find that the action 
is invariant under the following field redefinitions (cf. \refs\rmah)
\eqn\tr{
\eqalign{\chi^{'}_{4} &= {bd+ac\, e^{-2\phi_{4}}\over 
[(c\chi_{4}+d)^{2}+c^{2}e^{-2\phi_{4}}]^{1\over 4}}+
{ac\chi^{2}_{4}+(ad+bc)\chi_{4} \over [(c\chi_{4}+d)^{2}+
c^{2}e^{-2\phi_{4}}]^{1\over 4}
} \cr 
e^{-\phi'_{4}}&={e^{-\phi_{4}}\over [(c\chi_{4}+d)^{2}+
c^{2}e^{-2\phi_{4}}]^{1\over 4}} \cr
H^{(1)'}&= d\,H^{(1)}-c\,H^{(2)} \cr
H^{(2)'}&= [(c\chi_{4}+d)^{2}+c^{2}e^{-2\phi_{4}}]^{3\over 4}
(-b \,H^{(1)}+a\,H^{(2)}) \cr
G_{\mu\nu}^{'}&= [(c\chi_{4}+d)^{2}+c^{2}e^{-2\phi_{4}}]^{1\over 2} 
G_{\mu\nu}
}
}
Comparing with the corresponding transformations for the 
ten-dimensional fields
we find that the four-dimensional ones acquire an ``anomalous weight'' 
due to the fact
that $V_6$, as measured in the string frame, does transform under 
$SL(2,{\bf R})$. 

The algorithm to be described in the following uses as input diagonal vacuum 
solutions to 
Einstein equations with two {\it commuting} space-like Killing vectors. 
Infinite 
dimensional families of solutions of this kind 
are known \refs\rcchm \refs\rcchf \refs\rfi \refs\rverd. 
After transforming these diagonal vacuum solutions
to generate off-diagonal terms in the metric, we generalize them to include
a minimally coupled massless scalar field. A conformal rescaling of the
Einstein metric gives solutions to 
four-dimensional dilaton gravity and a T-duality 
transformation generates an antisymmetric rank-two 
tensor $B_{\mu\nu}^{(1)}$ from the off-diagonal components of the metric. 
This leaves us with string cosmology solutions of both the heterotic and
type-IIB superstring. In the latter case, we still can generate 
non-trivial background
R-R fields by performing a $SL(2,{\bf R})$ 
rotation of the solution (cf. \refs\rps). 
In the next section, we will detail the algorithm and in Section 3 a concrete 
application will be worked out. Finally, in Section 4 we will summarize 
our conclusions.

\newsec{String cosmology from General Relativity}

After having outlined the key steps to be followed in the 
construction of solutions,
we now turn to the generating algorithm itself. In order not to get too
involved with the details we shall ommit some technicalities, referring in
cases to the original papers. We present 
here only those details and expressions  
necessary to get the final solution and to keep 
the presentation self-consistent. 
As mentioned above our starting point is a globally 
diagonalizable $G_2$ line element 
which may be put into the following convenient form
\eqn\dosuno{
ds^2 = e^{f(t,z)}(-dt^2+dz^2)+ \gamma_{ab}(t,z)dx^{a}dx^{b}, 
\hskip 1cm a,b=1,2.
}
The local behavior of the spacetime is defined by the gradient of the
transitivity surface area $K(t,z)\equiv\sqrt{\det\gamma_{ab}}$,
which can be globally timelike, spacelike, null or vary from
point to point. We will not restrict our attention to any of these specific
cases keeping the determinant arbitrary; just mention that in cosmology
one is most interested in a globally timelike case, which includes
unisotropic Bianchi I-VII models and their inhomogeneous
generalizations, and a general case where the gradient of the transitivity
area may vary from point to point. The latter includes 
the most general Bianchi VIII
and IX types with local rotational symmetry with or without 
inhomogeneities \refs\rcchf, precisely the case which will serve us as 
example in this paper. If  one of the Killing vector fields 
$\xi_{1}=\partial_{1}$ and 
$\xi_{2}= \partial_{2}$ is orthogonal to the hypersurface obtained 
by dragging the surface $t$-$z$ along the other Killing vector 
field then the 
spacetime is globally diagonalizable. We thus start from 
such a spacetime in vacuum 
and notice that we can write 
\eqn\seed{
\gamma_{ab}(t,z)dx^a dx^b=K(t,z)\left[e^{p(t,z)}(dx^1)^2+ 
e^{-p(t,z)}(dx^2)^2\right],
}
where the function $p(t,z)$ satisfies the linear wave equation
$$
{d\over dt}\left(K\dot{p}\right) - {d\over dz}\left(K\,p^{'}\right)=0
$$
and the other metric function $f(t,z)$ may be found by a quadrature 
\refs\rcchm \refs\rcchf \refs\rverd.

One now applies one of the standard techniques to generate the non-diagonal
solution starting from the diagonal seed \seed. This can 
be done in multiple ways:
either using inverse scattering method of ref. \rbk, the solution generating 
algorithm described in \hfm, or an appropriately addapted HKX 
transformation \rkit. Actually, 
even a simple Ehlers rotation of the Killing vectors
\rehlers\ will do the job. Although this last procedure
does not generate a genuinely non-diagonal solution (one may globally
re-diagonalize the metric leading to a new solution),
combined with a T-duality transformation in the string frame it will 
produce the desired 
non-vanishing components for the $B_{\mu\nu}^{(1)}$ field.

Assuming that the non-diagonal vacuum solution has been constructed, 
we now have
to solve Einstein equations for the metric \dosuno\ with a minimally coupled 
{\it massless} scalar
field (our proto-dilaton). Fortunately, these solutions may be quite easily
obtained, since a massless scalar field has the same characteristics of 
propagation as
gravity, and does not introduce any extra degree of 
non-linearity into the problem. 
In fact, if $\phi(t,z)$ is the scalar field, the new solution of the 
Einstein equations 
may be written as \refs\rlwch
\eqn\g{
\eqalign{
\gamma_{ab}(t,z)&= \gamma_{ab}^{\rm vac}(t,z), \cr
f(t,z)&= f^{\rm vac}(t,z) + f^{\rm  sc}(t,z),
}
}
where $f^{\rm vac}(t,z)$ is the vacuum solution of the line element \dosuno, 
$f^{\rm sc}(t,z)$ is determined by
\eqn\f{
\eqalign{
\dot{f}^{\rm sc}&= {2K \over K^{'2}-\dot{K}^{2}}\left[2K^{'}\dot{\phi}\phi^{'}
-\dot{K}(\dot{\phi}^{2}+\phi^{'2})\right], \cr 
f^{'{\rm sc}}&={2K \over K^{'2}-
\dot{K}^{2}}\left[K^{'}(\dot{\phi}^{2}+\phi^{'2})
-2\dot{K}\dot{\phi}\phi^{'}\right]
}
}
and the scalar field $\phi(t,z)$ verifies the following linear 
differential equation
\eqn\above{
{d\over dt}(K\dot{\phi})-{d\over dz}(K\phi^{'})=0.
}
If the gradient of the transitivity surface area is globally 
timelike one may choose 
$K\sim  t$, and the solutions of the equation may be presented 
as combinations of the Bessel functions of the first and second kind \refs\rfi
$$
\eqalign{
\phi&=\beta\log{t}+ {\cal L}\{A_{\omega}\cos[\omega(z+z_0)]J_{0}(\omega t)\}+
{\cal L}\{B_{\omega}\cos[\omega(z+z_0)]N_{0}(\omega t)\} \cr
&-
\sum_{i} d_{i}{\rm arc}\cosh\left({z+z_{i}\over t}\right),
}
$$
where ${\cal L}$ indicates linear combinations of the terms in curly brackets, 
and $\omega$ can have a discrete or continous spectrum. 
On the other hand, in the more general case when the gradient varies from 
point to point 
and the spatial sections have $S^{3}$ topology, as it happens in the case 
of Bianchi 
IX models, $K\sim \sin{t} \sin{z}$ \refs\rgowdy\ and the general solution 
of the equation 
\above\ can be expanded in Legendre polinomials of the first and second kind
\eqn\dil{
\eqalign{
\phi&=\alpha_{1}\log\left|\tan{t\over 2}\right|+\alpha_{2}
\log\left|\tan{z\over 2}\right|
+\alpha_{3}\log|\sin{t}\sin{z}| \cr 
&+\sum_{\ell=0}^{\infty}\left[A_{\ell}P_{\ell}(\cos t)+B_{\ell}
Q_{\ell}(\cos t)\right]
\left[C_{\ell}P_{\ell}(\cos z)+D_{\ell}Q_{\ell}(\cos z)\right].
}
}

Equations \g, \f\ and \dil\ summarize the preliminary 
construction in the Einstein frame. 
Now we transform the solution into the string frame 
by the conformal transformation 
\rtt\rwitt
$$
ds^2 \longrightarrow e^{2\phi(t,z)} ds^2.
$$
This provides us with a solution to four-dimensional dilaton gravity. 
We are, nevertheless,
interested in having non-trivial values for other background fields, mainly the 
two-form potential $B_{\mu\nu}^{(1)}$ appearing in \univ. To accomplish 
this we can use a
T-duality transformation. Taking adapted coordinates in which 
$x^{0}$ denotes the coordinate along the Killing vector chosen to dualize, 
we find new values
for $(G_{\mu\nu},\phi,B_{\mu\nu}^{(1)})$ given by \refs\buscher
\eqn\bus{
\eqalign{{\tilde G}_{00}&={1\over G_{00}}, \hskip0.8cm 
{\tilde G}_{0\mu}={B_{0\mu}^{(1)}\over G_{00}},
\hskip0.8cm {\tilde G}_{\mu\nu}=G_{\mu\nu}-{G_{0\mu}G_{0\nu}+
B_{\mu0}^{(1)}B_{0\nu}^{(1)}\over G_{00}}, \cr
{\tilde B}_{0\mu}^{(1)} &= {G_{0\mu}\over G_{00}}, \hskip0.8cm 
{\tilde B}_{\mu\nu}^{(1)}=B_{\mu\nu}^{(1)}
-{G_{0\mu}B_{0\nu}^{(1)}+G_{0\nu}B_{\mu0}^{(1)}\over G_{00}}, \cr
{\tilde \phi}&=\phi-\log\sqrt{G_{00}}.
}
}
The new background fields so obtained automatically satisfy the field equations 
derived
from the effective action \univ\ \refs\buscher. Since we have 
two commuting Killing
vectors we can dualize with respect to both of them to end up with non-zero
values for several components of $B_{\mu\nu}^{(1)}$. At this point 
it is important to stress  
that we are using Buscher's formulae \bus\ just as a formal 
procedure to generate new 
solutions of the low-energy field equations. 

So far, the application of our algorithm leads to solutions to the low-energy
equations for the ``universal massless spectrum'' 
of the four dimensional heterotic and 
type-II superstring. In the latter case we would 
be interested in getting solutions
including also background values for the R-R fields. In the case of the 
type-IIB superstring
this can be done using the invariance of the whole effective action 
$S_{\rm univ}+S_{\rm md}^{\rm IIB}$ under the four-dimensional 
$SL(2,{\bf R})$ transformation
\tr\ \rps. In particular, starting with a solution in which 
$\chi=B_{\mu\nu}^{(2)}
=0$, we can generate non-trivial values for these R-R fields, 
thus completing our
generating algorithm. The important point here 
is to realize that, in this last step,
we are doing more than just performing a S-duality transformation. 
Although $SL(2,{\bf R})$
is a symmetry of the low-energy effective type-IIB supergravity, 
only a subgroup 
$SL(2,{\bf Z})$ is an actual symmetry of the full 
underlying string theory. Therefore
in perfoming a {\it generic} $SL(2,{\bf R})$ rotation we 
will end up with solutions 
which are not equivalent to the original one at 
the level of the string theory.

\newsec{Inhomogeneous Mixmaster string cosmologies}

To illustrate the use of the algorithm we suggest to 
look at the following example. It is well 
known in Cosmology that Bianchi type IX models 
(the famous Mixmaster universes
\refs\rmisner), let alone their inhomogeneous generalizations, 
represent a very interesting
class of models to study. Due to the presence of spatial curvature 
and the nontrivial $S^3$
topology one may investigate their effects on the initial expansion.
We consider here the locally rotationally symmetric (LRS) case which
may be incorporated within the Einstein-Rosen spacetimes \refs\rcchf. 
To shorten the
procedure we pass over the first two steps and start 
directly with the solution to 
Einstein-Klein-Gordon equations obtained
previously by one of us \refs\rcchf. We can write it in the string frame as
\eqn\uno{
\eqalign{
ds^2&=\left(\tan{t\over 2}\right)^{2M\over k}
e^{2\phi_2(t,\theta)}\left\{e^{f(t,\theta)}(-dt^2+d\theta^2)+[I_1^2(t)
\sin^2\theta+
I_3^2(t)\cos^2\theta]d\varphi^2  \right. \cr
&+ \left. I_3^2(t) d\psi^2+2\,I_3^2(t)\cos \theta\, d\varphi \,d\psi\right\}
}
}
where the functions $f(t,\theta)$, $I_1(t)$ and $I_3(t)$ are defined by
$$
\eqalign{
f(t,\theta)&=2\log I_1(t)+C^2\sin^2 2t (1+3\cos^2\theta) \sin^2 \theta
\cr
&+2\, C^2\sin^2t (1-3\cos^2t)\sin^4\theta-
16\,{M\over k}C\cos t\sin^2\theta, \cr
I_1^2(t)&= {k^2 \over 2A}{\left(\tan{t\over 2}\right)^{2A\over k}+
\left(\cot{t\over 2}\right)^{2A\over k}\over (\tan{t\over 2}+
\cot{t\over 2})^2},\cr
I_3^2(t)&={2A\over \left(\tan{t\over 2}\right)^{{2A\over k}}+
\left(\cot{t\over 2}\right)^{2A\over k}},
}
$$
and the coordinates $0\leq\theta\leq\pi$, $0\leq\phi\leq 2\pi$ and 
$0\leq\psi\leq4\pi$ 
are taken to be Euler angles. For the sake of simplicity, we have chosen 
a scalar dilaton 
field $\phi(t,\theta)=\phi_{0}(t)+\phi_{1}(t,\theta)$ 
containing just two modes, 
the homogeneous one 
$$
\phi_0(t)={M\over k}\log \tan {t\over 2},
$$
and the inhomogeneous growing mode $\phi_{1}(t,\theta)$ 
compatible with the boundary conditions 
at $\theta =0, \pi$ \refs\rcchf,
$$
\phi_{1}(t,\theta)={C\over 4} (1-3\cos^2t)(1-3\cos^2\theta).
$$
The $S^3$ topology plays an important r\^{o}le near the initial 
singularity by imposing strong
restrictions on the allowed modes of the scalar field, 
excluding for example strictly decreasing
inhomogeous solutions.

After T-dualizing with respect to the Killing vector 
$\xi=\partial_{\varphi}$ we
trade the off-diagonal term in the metric, $G_{\varphi\psi}$, 
for a non-vanishing component
$B_{\varphi\psi}^{(1)}$ of the rank-two tensor potential. The 
dual metric has finally the form
\eqn\dos{
\eqalign{
d\tilde{s}^2&=\left(\tan{t\over 2}\right)^{2{M\over k}}e^{2\phi_1(t,\theta)}
\left[e^{f(t,\theta)}(-dt^2+d\theta^2)+{I_1^2(t) I_3^2(t) \sin^2\theta\over
I_1^2(t)\sin^2\theta+I_3^2(t)\cos^2\theta}d\psi^2\right]
\cr
&+\left(\tan{t\over 2}\right)^{-2{M\over k}}{e^{-2\phi_{1}(t,\theta)}
\over I_1^2(t)\sin^2\theta+I_3^2(t)\cos^2\theta} d\varphi^2 ,
}
}
while the new dilaton field $\tilde{\phi}(t,\theta)$ turns out to be
\eqn\nd{
\tilde{\phi}=-{1\over 2}\log[I_1^2(t)\sin^2\theta+I_3^2(t)\cos^2\theta].
}
The newly generated two-form field can be expressed as 
$$
\tilde{B}(t,\theta)={I_3^2(t)\cos\theta\over I_1^2(t)\sin^2\theta+
I_3^2(t)\cos^2\theta}\,
d\varphi\wedge d\psi.
$$
It is remarkable that the expression \nd\ for the transformed dilaton 
is independent of
the particular dilaton we began with. However, 
the latter leaves its imprint on the geometry,
as can be seen from \dos.

Several interesting things happen with this solution. 
When the homogeneous part of the
dilaton $\phi_0$ is absent the solution is everywhere 
regular in the sense of curvature
invariants and fields, whatever the value of $k$, except for 
the points located along
the axis $\theta=0,\pi$ at $t=0$. This can be traced back to 
the fact that those 
points are on singular
orbits of the Killing vector field $\partial_{\varphi}$. 
When $\phi_{0}(t)\neq 0$
we get a Big-Bang singularity as $t$ goes to zero, 
as well as a Big-Crunch in
$t=\pi$. With respect to the homogeneous character of 
the initial singularity, we find that
for the metric before T-duality \uno\ the singularity is 
approached homogeneously,
whereas \dos\ gives rise to an inhomogeneous Big-Bang. 
As will be discussed later this 
is not unexpected after all, since some types of 
inhomogeneities are generated by the 
T-duality transformation.   
 
Suppose now the inhomogeneous part of the 
dilaton field is switched off, $C\sim 0$ and $\phi_{0}(t)\neq 0$. 
The solution before 
T-duality describes then a Bianchi IX LRS cosmology which 
becomes inhomogeneous after 
the transformation. Had we started with the isotropic 
Friedman-Robertson-Walker
solution ($2I_1^2=2I_3^2=k\sin t$) before T-dualizing, 
we would have finished with 
anisotropic Kantowski-Sachs type cosmology,
$$
\eqalign{
d\tilde{s}^{2}&=\left(\tan{t\over 2}\right)^{\pm\sqrt{3}}I_{1}^{2}(t)(
-dt^2+d\theta^2+\sin^{2}\theta d\psi^{2})\cr
&+{1\over I_1^2(t)}
 \left(\tan{t\over 2}\right)^{\mp\sqrt{3}}
d\varphi^{2},
}
$$
which falls into the class of solutions recently studied 
by Barrow and D\c{a}browski
in \refs\rsc.

One thing to be learned from the above examples is that inhomogenous 
and/or anisotropic
spacetimes can be related to homogeneous and 
isotropic backgrounds via T-duality. 
Since this is a symmetry of string
theory, and therefore strings cannot distinguish 
between dual spacetimes, it seems  
that isotropy and homogeneity, and even some types of 
curvature singularities,
become less intuitive concepts in string theory than 
they are in classical General Relativity. This is not 
really a surprise, since it is well 
known that string theory forces us to revise the 
physical r\^{o}le of other classical 
concepts, such as the topology of the spacetime or the 
value of the cosmological constant 
\refs\raabl. Imagine now an inhomogeneous spacetime 
T-dual to a homogeneous one, so the 
physics at the string scale is the same in both 
backgrounds\foot{It is important to stress
that physics in dual backgrounds looks different 
{\it only} when described using momentum 
modes in both of them. If one, however, uses the 
winding states associated with the compact
isometry in the dual background, the physical equivalence 
manifest itself explicitly.}. 
This seems to 
pose an apparent puzzle, for in the field theory limit 
($\alpha^{'}\rightarrow 0$) we are left
with point particles which could, in principle, 
probe these inhomogeneities. Looking 
more carefully at the problem it turns out that this 
is not really the case, because 
of the hidden presence of the Planck length 
$\sqrt{\alpha^{'}}$ in Buscher's formulae. This 
makes the typical scale of these 
T-duality-generated inhomogeneities to be of the 
order of the Planck 
length, so they cannot be detected using low-energy 
(particle) probes. Thus the dual
spacetime looks as homogeneous as the original one when 
looked at large distances.
The moral of the story is then that there are certain 
types of inhomogeneities, those
that can be removed by a T-duality transformation, which 
in a sense are ``pure gauge"
from the string theory point of view. In the same vein, 
this generalizes to the case when
we start with an inhomogeneous background: now we have, 
in addition to
the long-wavelength inhomogeneities inherited from the 
original geometry, those generated
by T-duality. Since their typical scale is of the order 
of the Planck length
they look like ripples on a smooth manifold. 

\newsec{Conclusions}

Before closing let us summarize the main results obtained. 
We have designed a
general algorithm which allows the construction of 
cosmological solutions to the
low-energy effective theory of the heterotic and 
type-IIB superstring, starting with
vacuum solutions to Einstein equations. This 
provides us with a powerful tool to
construct new solutions in string cosmology, given 
the huge amount of vacuum spacetimes 
with two or more commuting spacelike Killing vectors 
known in General Relativity. Starting 
from either homogeneous or inhomogeneous seed 
metrics, we pass to Einstein-Klein-Gordon 
solutions, to end up after a conformal rescaling 
with solutions to dilaton gravity. 
Using T-duality and the S-duality of the type-IIB 
superstring we finally get 
cosmological spacetimes for both the heterotic 
and type-IIB theories, in the latter 
case also including non-trivial R-R fields.

As an illustrative example we have constructed 
the string theory version of the
inhomogeneous Mixmaster-type cosmology obtained in 
ref. \refs\rcchf. We have also discussed 
the r\^{o}le played by inhomogeneities in string 
cosmology and argued that there is a class of
those, the ones generated by T-duality, that have 
no physical relevance at any energy scale.
Thus, T-duality classifies inhomogeneous spacetimes 
into physically equivalent pairs.
As it happens with many other issues, string theory 
seems to challenge our physical 
intuition about the real meaning of a homogeneous geometry.

\newsec{Acknowledgements}

We would like to thank A. Ach\'ucarro, J.L.F. Barb\'on, 
I.L. Egusquiza, J.L. Ma\~nes and
M.A. Valle-Basagoiti for interesting discussions and 
suggestions. This work has been 
supported in part by a Spanish Ministry of 
Education Grant (CICyT) PB93-0507 and
a Basque Country University Grant UPV/EHU/72.310EBO36/95. 
R.L. and M.A.V.-M. acknowledge financial support 
from the Basque Goverment 
under Fellowships BFI94-094 and BFI96-071 respectively.

\listrefs

%\vfill\eject
\bye